\documentclass[twocolumn]{aastex631}
\usepackage[T1]{fontenc}
\usepackage{amsmath}
\usepackage{amssymb}	
\usepackage{graphicx,aas_macros}
\usepackage{multirow}
\usepackage{subfigure}
\usepackage{color}
\newcommand{\beq}{\begin{equation}}
\newcommand{\eeq}{\end{equation}}
\newcommand{\beqa}{\begin{eqnarray}}
\newcommand{\eeqa}{\end{eqnarray}}

\def\stacksymbols #1#2#3#4{\def\theguybelow{#2}
        \def\verticalposition{\lower#3pt}
        \def\spacingwithinsymbol{\baselineskip0pt\lineskip#4pt}
        \mathrel{\mathpalette\intermediary#1}}
\def\intermediary #1#2{\verticalposition\vbox{\spacingwithinsymbol
        \everycr={}\tabskip0pt
        \halign{$\mathsurround0pt#1\hfil##\hfil$\crcr#2\crcr
                \theguybelow\crcr}}}

\def\gta{\stacksymbols{>}{\sim}{3}{.5}}

\shorttitle{
AGN Feedback, X‑ray and Dusty CO Clouds in Group Cores }
\shortauthors{Temi et al.}

\begin{document}
\title{
AGN Feedback and the Development of Dusty Multiphase Gas in X-ray Emitting Elliptical Galaxies\\
}

\correspondingauthor{P.~Temi}
\email{pasquale.temi@nasa.gov}

\author[0000-0002-8341-342X]{Pasquale Temi}
\affiliation{Astrophysics Branch, NASA - Ames Research Center, MS 245-6, Moffett Field, CA 94035}

\author[0000-0001-5338-4472]{Francesco Ubertosi}
\affiliation{Dipartimento di Fisica e Astronomia, Universita` di Bologna, via Gobetti 93/2, I-40129 Bologna, Italy}
\affiliation{Istituto Nazionale di Astrofisica - Istituto di Radioastronomia (IRA), via Gobetti 101, I-40129 Bologna, Italy}

\author[0000-0001-9807-8479]{Fabrizio Brighenti}
\affiliation{Dipartimento di Fisica e Astronomia, Universita` di Bologna, via Gobetti 93/2, I-40129 Bologna, Italy}
\affiliation{University of California Observatories/Lick Observatory, Department of Astronomy and Astrophysics, Santa Cruz, CA 95064, USA}

\author[0000-0003-2552-3871]{Alexandros Maragkoudakis}
\affiliation{Astrophysics Branch, NASA - Ames Research Center, MS 245-6, Moffett Field, CA 94035}
\affiliation{Bay Area Environmental Research Institute, Moffett Field, California 94035, USA}

\author[0000-0001-6638-4324]{Valeria Olivares}
\affiliation{Departamento de F\'isica, Universidad de Santiago de Chile, Av. Victor Jara 3659, Santiago 9170124, Chile}
\affiliation{Center for Interdisciplinary Research in Astrophysics and Space Exploration (CIRAS), Universidad de Santiago de Chile, Santiago 9170124, Chile}

\author[0000-0002-2212-5395]{Alexandre Amblard}
\affiliation{NASA Ames Research Center - Aviation Systems Division}
\affiliation{Crown Consulting, Inc Arlington, VA}

\author[0000-0003-2754-9258]{Massimo Gaspari}
\affiliation{Department of Physics, Informatics and Mathematics, University of Modena and Reggio Emilia, 41125 Modena, Italy}

\author[0000-0002-0843-3009]{Myriam Gitti}
\affiliation{Dipartimento di Fisica e Astronomia, Universita` di Bologna, via Gobetti 93/2, I-40129 Bologna, Italy}
\affiliation{Istituto Nazionale di Astrofisica - Istituto di Radioastronomia (IRA), via Gobetti 101, I-40129 Bologna, Italy}

\author{Pamela M. Marcum}
\affiliation{Astrophysics Branch, NASA - Ames Research Center, MS 245-6, Moffett Field, CA 94035}

\author[0000-0002-2691-2476]{Kevin Fogarty}
\affiliation{Astrophysics Branch, NASA - Ames Research Center, MS 245-6, Moffett Field, CA 94035}

\author[0000-0003-3249-4431]{Alejandro S. Borlaff}
\affiliation{Astrophysics Branch, NASA - Ames Research Center, MS 245-6, Moffett Field, CA 94035}
\affiliation{Bay Area Environmental Research Institute, Moffett Field, California 94035, USA}

\author{William G. Mathews {}\textsuperscript{\dag}}
\renewcommand{\thefootnote}{\dag}
\footnotetext{Deceased.}
\renewcommand{\thefootnote}{\arabic{footnote}}

\begin{abstract}

This paper investigates the physical and kinematic properties of dust-rich regions in a small sample of group-centered elliptical galaxies, emphasizing their connection with the hot X-ray emitting gas and detailed dust grain characteristics. Comprehensive multi-wavelength data—including H$\alpha$ and CO emission detected by MUSE and ALMA—demonstrate the presence of dust clouds embedded within complex, hot X-ray atmospheres shaped by AGN feedback. 
X-ray images show bubbles and cavities surrounded by bright rims. We find that dust regions containing molecular gas traced by CO are preferentially located at the rims of these X-ray cavities, suggesting that AGN-driven outflows enhance the condensation of cold, dusty gas at these compressive interfaces. Kinematic measurements indicate that molecular and ionized gas phases are dynamically and spatially linked, supporting the framework of a multiphase medium arising from the top-down condensation rain in the hot plasma and related chaotic cold accretion. 
Crucially, spatial variations in the total-to-selective extinction ratio $R_V$ show that regions where dust, CO, and H$\alpha$ emission coincide exhibit notably smaller $R_V$ values, implying steeper extinction curves and the predominance of smaller or less evolved dust grains within these mixed-phase environments. 
This contrasts with larger $R_V$ values found elsewhere in the dust clouds, suggesting grain growth or survival mechanisms within shielded cold gas.

\end{abstract}

\section{Introduction }
\noindent
In recent years, multi-wavelength observations of massive elliptical galaxies have revealed the presence of a complex and multiphase interstellar medium (ISM) to an unprecedented level of detail \citep[e.g.][]{temi07a,temi07b,werner18,Eskenasy2024, Olivares2019,temi22,Olivares2025}. 
Given their old stellar populations \citep{Trager00,Annibali06, Diniz17}, smooth stellar light and interstellar gas distributions \citep[e.g.,][]{Ziegler, Van05, Daddi05,Lagos2022,Loubser2022,osullivan17,olivares2022a}, and negligible star formation rates ($\ll$1\, M$_\odot$~yr$^{-1}$; \citet{werner14}), massive elliptical galaxies, and in particular in local brightest group galaxies (BGGs), serve as an ideal testbed for detailed investigations of dust-gas-radiation interactions in absence of confounding factors. 
The energetic central environment of BGGs, dominated by hot X-ray gas, AGN feedback, and turbulence, makes conditions inhospitable for the formation and sustainability of cold molecular clouds. Curiously, observations show molecular clouds and dust persist within a few central kiloparsecs, indicating localized survival in otherwise hostile conditions.

Molecular gas in the centers of BGGs has been both inferred and directly observed through [CII] and CO observations, respectively. \citet{crawford85} and \citet{werner14} argue that [CII] emission in massive ellipticals is an excellent tracer of cold molecular gas and that [CII] luminosities indicate the presence of large reservoirs of cold gas. Extended [CII] emission at 158$\mu$m has been observed in several local BGGs using the Herschel PACS spectrometer and the FIFI-LS IFU onboard SOFIA \citep{werner14,temi22}. The [CII] emission, which traces cold gas in neutral, molecular, or ionized form, is spatially coincident with optical H$\alpha+$[NII] emission, suggesting a common origin. 

Recent ALMA observations of few BGGs have  detected a number of CO-emitting clouds  distributed within a few kpc from the center of the X-ray emitting atmosphere. The ALMA CO features consist of numerous self-gravitationally unbound clouds apparently orbiting within few kpc from the galaxy center. These molecular clouds, totaling $\sim6.1 \times 10^7$ $M_{\odot}$, $\sim2 \times 10^6$ $M_{\odot}$, and $\sim2.6 \times 10^5$ $M_{\odot}$ were detected with ALMA in NGC~5044, NGC~5846, and NGC~4636 \citep{david14, temi18}. These galaxies have a vigorously active supermassive black hole (SMBH) that feeds accretion energy back into the hot atmosphere -- an extreme environment not normally thought to create or harbor molecular clouds. Evidently, these clouds were either ejected by the active galactic nucleus (AGN) jets from regions near the central SMBH or, more likely, cooled directly from the diffuse hot atmosphere.  

In the theoretical framework of Chaotic Cold Accretion (CCA; \citealt{gaspari13,Gaspari20}), turbulence and bulk motion in the hot halo triggers thermal instability, leading gas to condense into multiphase filaments and clumps, which intermittently feed the central BH. This mechanism can naturally account for patchy multiphase morphologies, chaotic dynamics, and thermo-kinematic correlations. Moreover, dust grain processing in the mixing layers between hot and cold phases may produce local variations in extinction curves, which is part of the focus of this study.

\begin{table*}[ht]
\begin{center}
\caption{Summary of the galaxy sample properties} 
\begin{tabular}{cclccccccc}
\hline
Galaxy & cz $^{(a)}$& type $^{(b)}$& $D$ $^{(a)}$ & scale & ${F_{{H\alpha}}}^{(c)}$ & $\log(M_{\mathrm{HI}})^{(d)}$ & $M(\mathrm{H_2})^{(d)}$ & $\log(M_{\mathrm{dust}})^{(e)}$ & $L_{\mathrm{X}}\ ^{(f)}$ \\
       & (km s$^{-1}$) & & (Mpc) & ($^{\prime\prime}$/kpc) & ($10^{-13}$ erg cm$^{-2}$ s$^{-1}$) 
       & (M$_\odot$) & ($10^{8}$ M$_\odot$) & (M$_\odot$) & ($10^{40}$~erg s$^{-1}$) \\
\hline
NGC~4472 & 981  & -4.8 & 16.1 & 12.8 & $<1.7^{*}$ 
         & <8.0 & <0.3 & 4.01$\pm$0.12 & 18.36$\pm$0.08  \\
NGC~4636 & 938  & -4.8 & 15.9 & 12.9 & $1.5\pm0.2$ 
         & 8.84 & 0.010$\pm$0.003 & 4.36$\pm$0.07 & 19.68$\pm$0.09  \\
NGC~5044 & 2782 & -4.8 & 35.7 &  5.8 & $1.4\pm0.3$ 
         & \nodata & 0.5 & 5.53$\pm$0.10 & 204.11$\pm$0.47 \\
NGC~5813 & 1956 & -4.9 & 29.2 &  7.0 & $0.52\pm0.09$ 
         & <7.94 & <0.11 & 4.01$\pm$0.18 & 43.88$\pm$0.10  \\
NGC~5846 & 1712 & -4.8 & 27.1 &  7.6 & $2.6\pm0.2$ 
         & 8.48 & 0.14$\pm$0.06 & 4.83$\pm$0.09 & 15.72$\pm$0.11 \\
\hline
\end{tabular}
\label{tbl:summary}
\end{center}
{\footnotesize
\vspace{1ex}
\noindent \textbf{Notes.} Column descriptions: 
(1) \textit{Galaxy}: galaxy name; 
(2) \textit{cz}: recession velocity in km/s; 
(3) \textit{type}: galaxy morphological type with T-type; 
(4) \textit{D}: distance in Mpc; 
(5) \textit{scale}: angular scale in arcseconds per kiloparsec; 
(6) \textit{${F_{{H\alpha}}}$}: H$\alpha$ flux in $10^{-13}$ erg cm$^{-2}$ s$^{-1}$ within a $20^{\prime\prime}$ aperture; 
(7) $\log(M_{\mathrm{HI}})$: neutral hydrogen mass; 
(8) $M(\mathrm{H_2})$: molecular gas mass in units of $10^{8}$ M$_\odot$; 
(9) $\log(M_{\mathrm{dust}})$: dust mass; 
(10) $L_{\mathrm{X}}$: X-ray luminosity within $5r_{e}$ (the effective radius) in the 0.5 - 2 keV band. 
$^{*}$ Upper limit in H$\alpha$+[NII] emission based on rms = 2.1 dex \citep{gavazzi18}.
$^{(a)}$ taken from NASA/IPAC Extragalactic Database (NED); $^{(b)}$ taken from LEDA; $^{(c)}$ Observed with MUSE;
$^{(d)}$ taken from \citet{OSullivan2018}; $^{(e)}$ taken from \citet{amblard14}; $^{(f)}$ taken from \citet{babyk18}. 
} 
\end{table*}

H$\alpha$ emission demonstrates the central regions to indeed be energetic. In addition to their hot ($T\sim10^7$K) X-ray group-scale atmospheres \citep{Mathews03,Sun12,Goulding16, Anderson15,o'sullivan17}, about 50\% of the X-ray and optical bright massive ellipticals contain extended warm ($T\sim10^4$K) gas within 3-10 kpc from the center visible in H$\alpha$ line emission \citep[e.g.,][]{caon00,mcdonald11,sarzi13,werner14, lak18,OSullivan2018,Olivares2019}. Further evidence of energetic instability is indicated by erratic surface brightness distributions and velocities of H$\alpha$ gas, neither of which match those of the underlying stellar system \citep[e.g.,][]{olivares2022a}.

Finally, optical evidence of these centrally-located clouds is given by high-resolution imaging from the {\it Hubble Space Telescope}, which frequently reveals dust-rich clouds or disks within the central few hundred parsecs of elliptical galaxies \citep{dokkum95, lauer05}. These structures often exhibit morphological signs of transient disturbances, which are likely associated with AGN outbursts \citep{temi07a, temi07b}. Typical optically estimated dust masses in these central clouds are of the order of $\sim10^5$ $M_{\odot}$. \citet{Mathews03} showed that dusty gas ejected from red giant stars within about 1\,kpc of the galaxy centers can settle into the core while retaining a significant fraction of its original dust. Thus,
investigating the extinction properties and grain size distributions of dust clouds in the nuclear regions of BGGs offers valuable insights into their formation mechanisms, including condensation in stellar outflows, grain growth within molecular clouds, and destruction by shocks. 

The objective of this paper is to investigate how the spatial distribution of molecular gas and dust, and their properties, correlate with the emission and kinematics of energetic gas phases (X-ray and H$\alpha$) that are likely regulated by AGN feedback, in order to constrain the origin of dusty molecular clouds in a small sample of otherwise normal, massive, group-central elliptical galaxies.
Understanding the origin of molecular gas in BGGs is relevant because it reveals how massive, seemingly quenched ellipticals still acquire the cold, dusty fuel needed to power their central AGN and any residual star formation. In particular, distinguishing whether this gas arises from cooling of the hot X-ray halo, gas-rich mergers, or stellar mass loss allows us to connect the morphology and kinematics of molecular and dusty clouds with those of the X-ray and H$\alpha$-emitting phases governed by AGN feedback. This directly tests the self-regulated AGN feeding/feedback cycle, in which hot gas cools into molecular clouds, fuels the black hole, and is then heated, displaced, or destroyed by jets and outflows, thereby maintaining quenching in massive group-centered ellipticals.

 We present the sample and observations in  \S \ref{sample} and \S \ref{observations}. The results are delivered in \S \ref{results}, while the  correlations and properties of the molecular clouds relative to dust features and warm gas emission are discussed in \S \ref{discussion}. Finally, the conclusions are summarized in \S \ref{conclusions}.

\section{The galaxy sample}\label{sample}
In this study, we focus our analysis on a small set of group-centered elliptical galaxies.
The sample contains 5 of the brightest galaxies in X-ray bright
groups, where both observations and theory indicate significant hot gas
cooling. These galaxies have the most complete observational coverage, with
deep {\it Chandra} X-ray data, Multi Unit Spectroscopic Explorer (MUSE) H$\alpha$ observations, Spitzer IR data and detailed Hubble Space Telescope (HST) maps. 
Four galaxies (NGC 4636, NGC 5044, NGC 5813, and NGC 5846) have been observed with ALMA in the CO (2–1) emission line. The data from these observations have been reduced and analyzed, and the results have been presented in recent publications \citep{david14, temi18}.
A summary of the galaxy sample properties is shown in Table~\ref{tbl:summary}.

\section{Observations}\label{observations}

\begin{table}[ht!]
\begin{center}
 \caption{HST filters used in the analysis.}
\begin{tabular}{llc}
\hline\\
Name & Blue Filter& Red Filter\\
     NGC      & Name,  $\lambda$               & Name, $\lambda$\\
\hline\\
4472 & F555w, 5410\AA, V-band& F814w, 8353\AA, I-band \\
4636 & F547m, 5475\AA, Stromgren y& F814w, 8353\AA, I-band  \\
5044& \nodata & F814w, 8353\AA, I-band  \\
5813 & F555w, 5410\AA, V-band& F814w, 8353\AA, I-band  \\
5846 & F555w, 5410\AA, V-band& F814w, 8353\AA, I-band  \\
\hline\\
\end{tabular}
  \label{tbl:hst}
 \end{center}
\end{table}

\subsection{HST Data}
All the galaxies in the sample have been observed by the Hubble Space Telescope (HST) in recent years, utilizing various cameras and observing modes. From the available data in the HST archive, we selected a set of observations that are as uniform as possible in terms of science instruments and observing configurations. This selection ensures a more robust and consistent analysis of results across the target galaxies. 

The data were retrieved from the Hubble Legacy Archive and consist of two filters for each galaxy, except for NGC~5044, corresponding approximately to the V and I bands (see Table~\ref{tbl:hst} for details). 

\begin{figure*}[ht]
\begin{center}
\hskip-0.0cm
  \includegraphics[width=18.5cm]{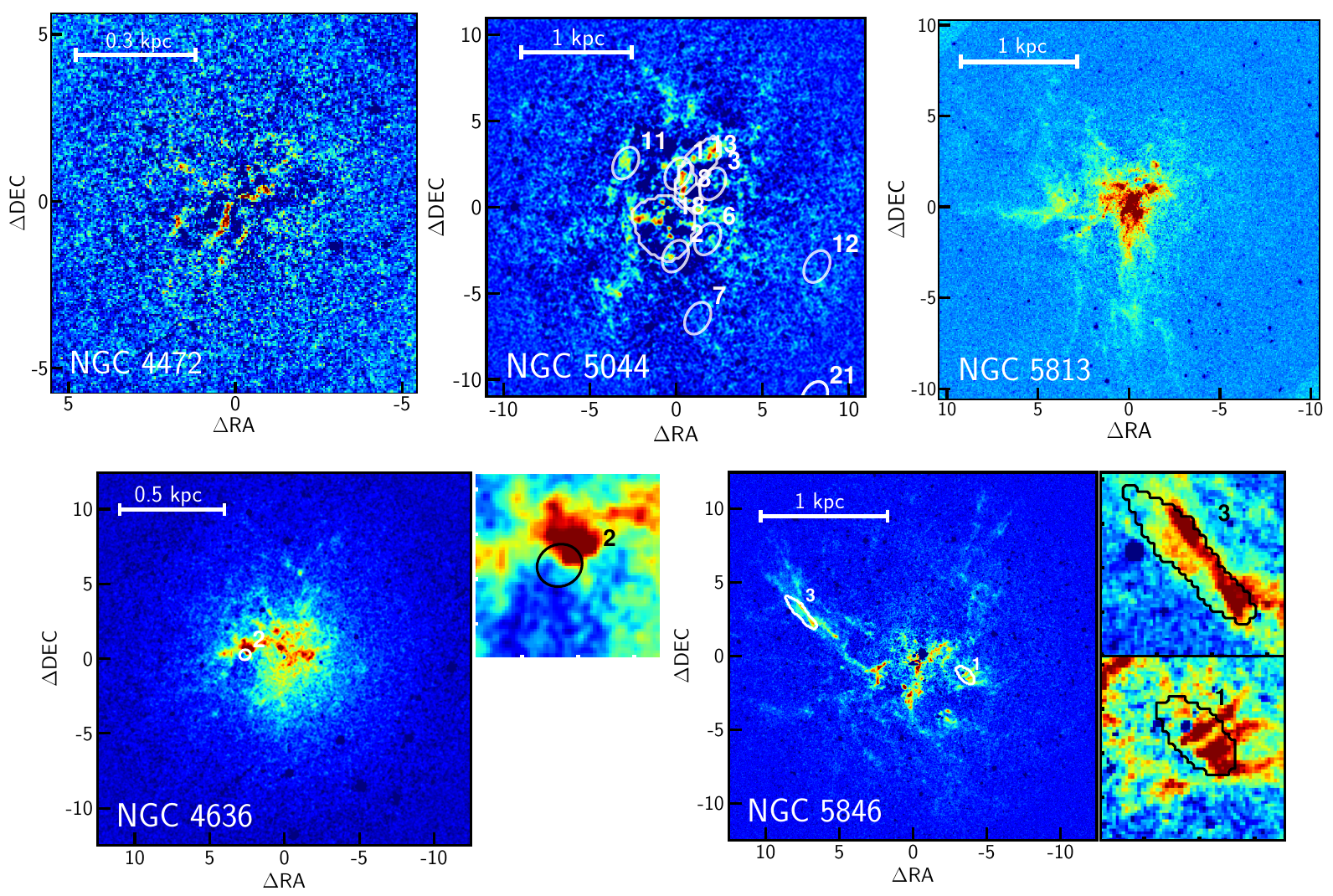}
  \caption{
HST images of the central few kpc of our sample galaxies. The optical dust maps were generated from archival HST data recorded with the Wide Field Planetary Camera 2 (WFPC2) using the broadband filters. 
Dust absorption features are evident as bright filamentary structures in the maps with increasing dust following the blue $\rightarrow$ green $\rightarrow$ yellow
$\rightarrow$ red color scheme. ALMA detection of CO clouds are indicated with white contours (black contours in the inset of NGC~4636 and NGC~5846) defined as the area where the emission
  line signal-to-noise is greater than 4.
}
\label{fg:dust1}
\end{center}
\end{figure*}

\begin{figure*}[ht]
\begin{center}
\hskip-0.0cm
  \includegraphics[width=18.5cm]{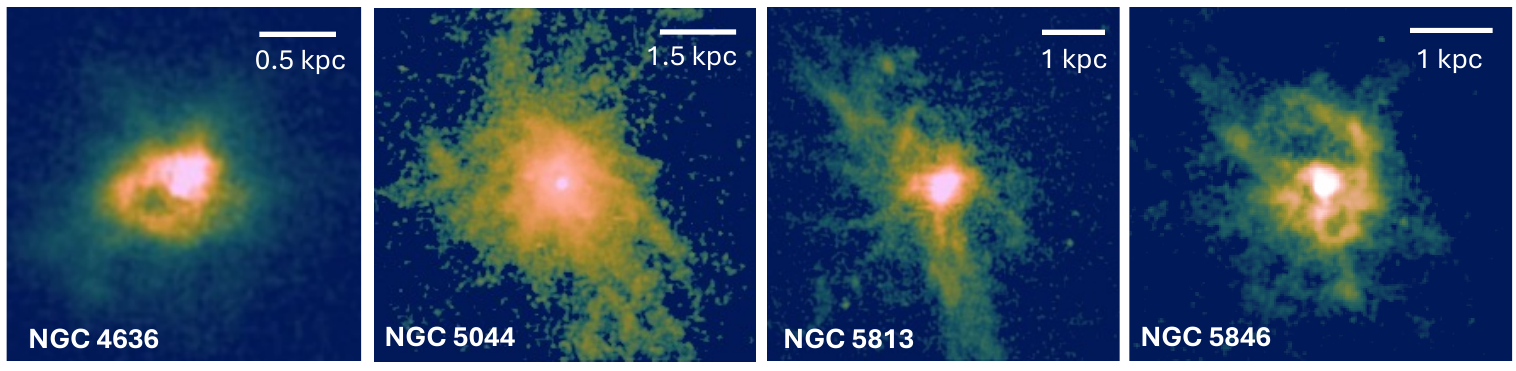}
  \caption{
MUSE H$\alpha$ maps of NGC~4636, NGC~5044, NGC~5813, and NGC~5846.
}
\label{fg:Ha_maps}
\end{center}
\end{figure*}

\subsection{MUSE Data}
MUSE optical integral field unit (IFU) observations were collected from the ESO Science Archive, which offers reduced, pipeline-processed science data products.  The MUSE observations of NGC~4636, NGC~5044, and NGC~5846 were obtained in the Wide Field Mode (WFM) with a mean resolution of R = 3000, covering a wavelength range between 4750\AA--9350\AA, and a sampling of 0.2" per pixel. 
Observations of NGC~5813 were not available in the public archive, as they were obtained during the instrument’s commissioning phase. The observations and data reduction are described in \citet{krajnovic15}, and the reduced data cube was kindly provided by the first author.\\
NGC 4472 has not been observed with MUSE, but shows no significant integrated H$\alpha$ emission in narrowband imaging surveys \citep{gavazzi18}. As reported in Table 1, only an upper limit on its total H$\alpha$+[NII] flux is found, implying extremely weak or negligible ionized gas.

\subsection{ALMA Data}

 We include recent ALMA CO(2–1) observations for three galaxies in our sample: NGC~4636 and NGC~5846 from Cycle 3 (project code 2015.1.00860.S; PI: Temi) and NGC~5044 from Cycle 0 (project code 2011.0.00735.S; PI: Lim). The array configuration, with baselines ranging from about 15 m to 640 m, delivered an angular resolution of about 0.6$^{\prime \prime}$ and a maximum recoverable scale of 5.4$^{\prime \prime}$. All observations were carried out in ALMA Band 6, with one spectral window centered on the CO(2–1) line at 0.5 MHz ($\sim$ 0.6 ${\rm km\,s^{-1}}$) resolution over a 937.5 MHz ($\sim$ 1200 ${\rm km\,s^{-1}}$) bandwidth. 
The observing setup, data reduction, and analysis are described in detail by \citet{temi18}. Although the observations were obtained at different epochs, all galaxies were processed in a uniform and consistent way: the same reduction tools and parameter settings were applied, and the data were calibrated and imaged with the CASA software package (version 6.5.1; CASA Team et al. 2022).

ALMA CO observations of NGC~5813 from Cycle 3 (project code 2015.1.00860.S; PI: Randall) are also present in the public archive. The data taken in Band 6 were  reduced and analyzed following the same procedures used for the other sample galaxies. ALMA CO(2–1) mapping of NGC~5813 shows non-detections of CO(2–1), with 3$\sigma$ upper limit on the integrated line flux of order of $\sim 0.3 \ Jy \ km \ s^{-1}$ over line widths of a $500 \ km \ s^{-1}$

Diffuse molecular gas has been detected in our galaxy sample using ALMA ACA and IRAM observations. 
In NGC5044, there is clear evidence of diffuse CO emission on kiloparsec scales, with most of the cold molecular gas concentrated within the central few kiloparsecs, and emission extending up to at least a 30$^{\prime \prime}$ ($\sim$ 5kpc) region \citep{schellenberger20, schellenberger21}. 

\subsection{Chandra Data}
All the galaxies in the sample have X-ray bright atmospheres of hot gas (see last column of Table~\ref{tbl:summary}), and as such, have been extensively targeted by X-ray telescopes. As we are interested in comparing the morphology and properties of warm gas and dust with the morphology of the hot gas at high ($\sim$arcsec) resolution, we consider the existing {\it Chandra} observations of the five galaxies. Specifically, we include {\it Chandra} ACIS observations of NGC~4636 (ObsID 323, 324, 3926, 4415, 209~ks in total), NGC~5846 (ObsID 788, 7923, 120~ks in total), NGC~5813 (ObsID 5907, 9517, 12951, 12952, 12953, 13246, 13247, 13253, 13255, 638~ks in total) and NGC~5044 (ObsID 798, 9399, 17195, 17196, 17653, 17654, 17666, 419~ks in total). 
Although available, we do not present X‑ray data for NGC 4472. In the absence of complementary MUSE $H_\alpha$ measurements and ALMA CO observations, the X‑ray information alone would add little to our comparative study and could be misleading.

\section{Results}\label{results}
\subsection{Dust Absorption Maps}
To produce maps of the flux absorbed by the dust in each filter, we
used the ELLIPSE function in the IRAF STSDAS
library\footnote{http://www.stsci.edu/institute/software\_hardware/stsdas;\\
http://iraf.noao.edu/iraf-homepage.html}
and a similar function in the Python library
photutils\footnote{https://photutils.readthedocs.io/en/stable/}\citep{bradley19}. These
codes calculate the underlying elliptical isophotes of the stellar
emission, and from these isophotes they deduce a model of the stellar
emission. The accuracy of these models is limited at the center of
galaxies due to the small number of pixels along the isophotes and at
larger radii where the signal-to-noise is low. Alternatively, when 2 HST filters
are available we approximate the I band filter as the model for the stellar emission
and fit a global stellar color between the I and V band filter. On one hand this method
reduces the dust absorption estimate by assuming no dust absorption in the I band and is
sensitive to stellar color gradient, on the other hand it can model any shape/geometry of stellar
emission. Of the five galaxies in the sample, the dust absorption maps for NGC 4472 and NGC 5044 were produced using the elliptical isophotes method.

\begin{figure*}[ht]
\begin{center}
\hskip-0.0cm
  \includegraphics[width=18.5cm]{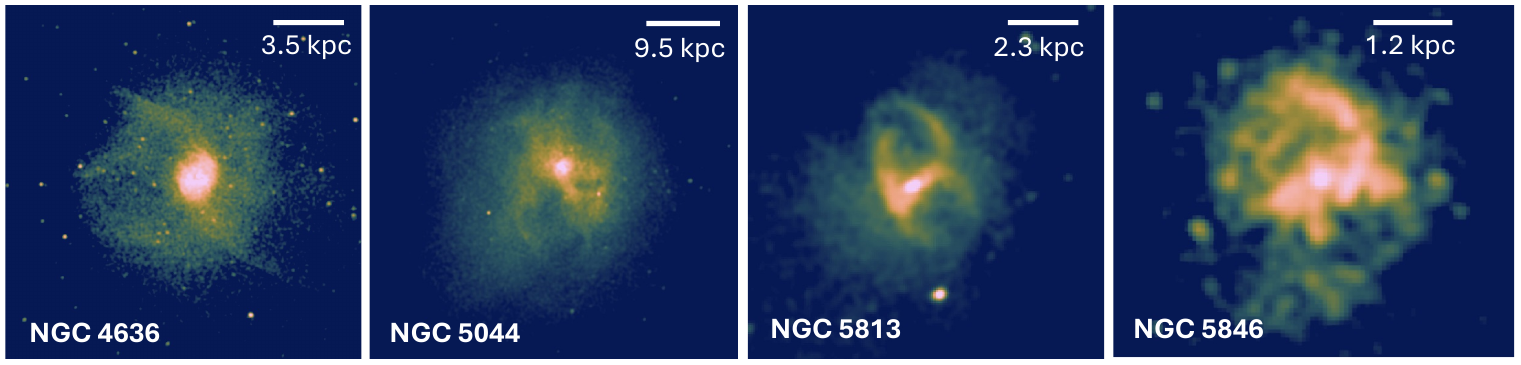}
  \caption{
Chandra X-ray maps in the 0.5 - 7 keV band of NGC~4636, NGC~5044, NGC~5813, and NGC~5846.
}
\label{fg:X_ray_maps}
\end{center}
\end{figure*}

In order to calculate the dust extinction 
we used the V--band HST map and one of the single-band model to detect regions of large
absorption. We set a threshold of 0.03 count/s on the difference between images and models that were smoothed with a 0.35\arcsec FWHM (to remove
isolate noisy pixels). Finally we clustered the selected pixels into area
of at least 50 pixels and obtain several high absorption clusters for each galaxy.
For each filter, we calculate the absorbed ($F_{abs}$) and unabsorbed ($F_{unabs}$) fluxes for each cluster
and computed the extinction
$A_{\lambda }=-2.5~log(F_{abs}/F_{unabs})$ in V and I--band (A$_I$ and A$_V$). The uncertainties on these magnitudes
are calculated using the statistical photon noise of the dust absorption image and
an estimate of the systematic uncertainty of the single-band model. The latter uncertainty
is estimated by calculating the root mean square of the dust absorption estimate where
its amplitude is very low (less than 0.01 count/s).

Figure~\ref{fg:dust1} shows color-scale HST images of dust absorption in the central few kpc
of our sample of galaxies. 
For the four galaxies observed with ALMA,
CO clouds, indicated with white contours, are defined as the area where the emission
  line signal-to-noise is greater than 4.
Small, chaotically arranged dusty fragments and filamentary structures
are visible against the stellar background in all the galaxies. 

\subsection{H$\alpha$ maps}

The MUSE spectral maps were modeled and fitted using the latest version of the Penalized PiXel-Fitting (\texttt{ppxf}) code \citep{Cappellari2017, Cappellari2023}. Continuum modeling and emission line fitting was applied to each individual pixel of the MUSE cubes. Although the MUSE wavelength coverage extends to 9300~\AA, we have truncated the spectra to 7400~\AA \ where the spectrum is less contaminated by sky residuals, while all the important optical emission (and absorption) lines remain. For the stellar continuum and stellar kinematics modeling the "E-MILES" stellar population synthesis templates \citep{Vazdekis2016} where used. Sets of multiple Gaussian components, in combination with the stellar templates, were used for fitting the optical emission lines and extracting the gas kinematics components. 

Maps of H$\alpha$ emission (see Figure \ref{fg:Ha_maps}) and the kinematics of the optical line-emitting gas are available for all galaxies in our sample except NGC~4472. In all maps, the line emission is spatially extended, with peaks consistently coincident with the optical galactic nucleus. Filamentary structures and bright compact knots are clearly visible within the central few kiloparsecs.

\subsection{X-Ray maps}

We obtained the publicly available \textit{Chandra} observations for our sample from the Chandra Data Archive\footnote{https://cda.harvard.edu/chaser/.}.
The data was reduced using CIAO \citep[version 4.16][]{fruscione06}.
All the data were reprocessed using the standard $chandra\_repro$ tool. Periods of strong background flares were filtered using the $lc\_clean$ script, and the threshold was set to match the blank-sky background maps. Point sources were detected using the CIAO task $wavedetect$ and later verified by visual inspection of the X-ray images. In order to merge the multiple ObsIDs of the same galaxy, the observations were first matched to the astrometry of the longest one by comparing the position of point sources in the images (using the $wcs\_match$ and $wcs\_update$ tools). We then created merged, exposure-corrected, background subtracted {\it Chandra} images in the 0.5-7 keV band with the $merge\_obs$ tool, that reprojects and combines multiple ObsIDs. These mosaic images are presented in Figure \ref{fg:X_ray_maps}. \\

The detailed kinematical analysis of the cold and warm gas phases, along with the multi-wavelength correlations (dust vs. $H_\alpha$ vs. X-ray distributions and cloud properties), are deferred to  \S \ref{discussion} (Analysis).
The analysis is restricted to the galaxies and dust clouds appearing in Tables \ref{tbl:masshalpha_part2} and \ref{tbl:masshalpha} based on data quality and completeness criteria.

\begin{figure}[ht]
\hspace{-0.5cm}
   \includegraphics[width=9.4cm]{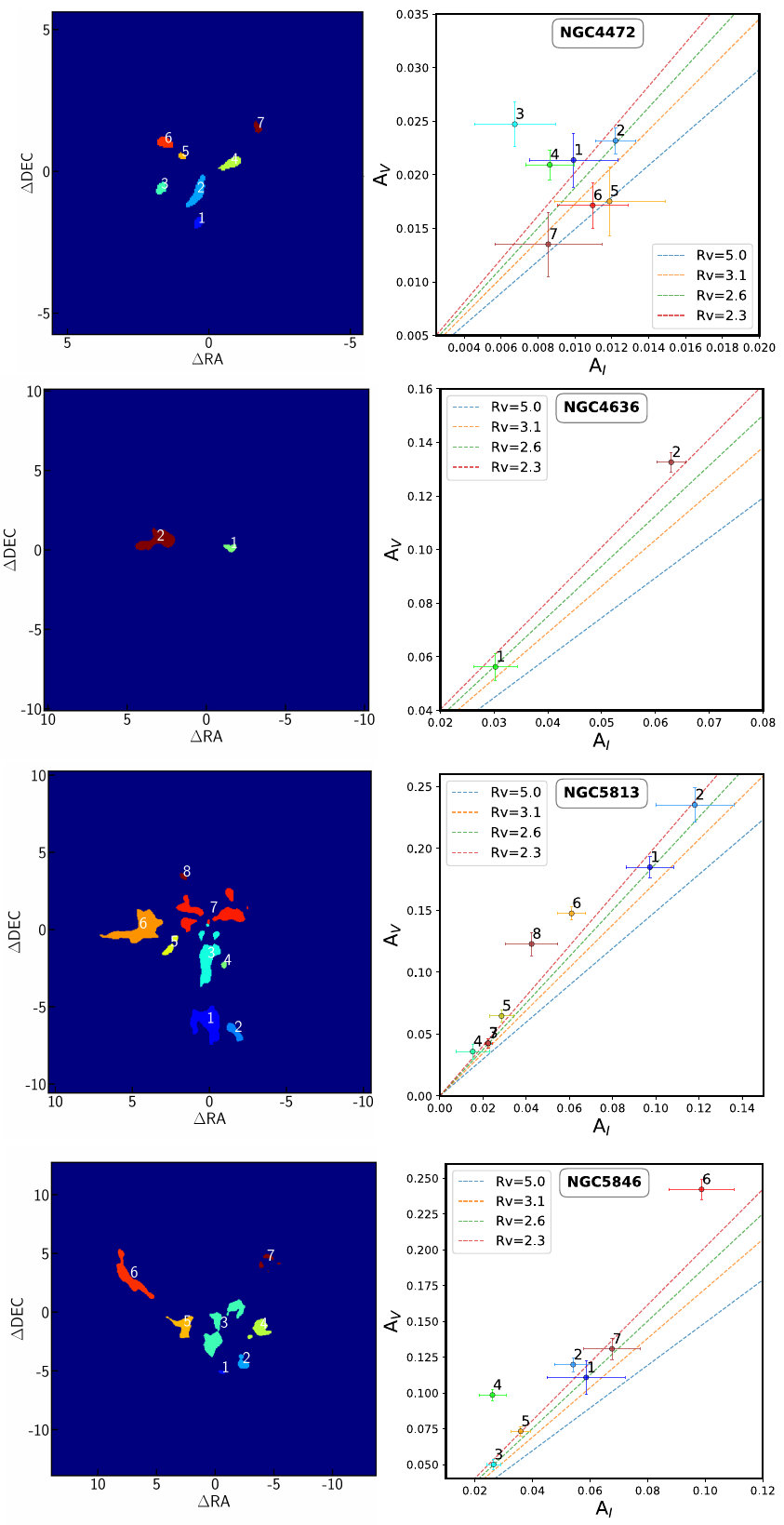} 
  \caption{
Left panels show the regions of strongest dust absorption where we measure the extinction; colors and numerical labels mark individual regions in each galaxy. The right panels display the corresponding A$_V$ and A$_I$ values, using the same colors for each region, together with their total uncertainties.
}
\label{fg:ratio}
\end{figure}

\subsection{ALMA maps}
Detailed CO maps and kinematic analyses are presented in \citet{temi18}; here we provide a brief summary of their findings.
ALMA CO(2–1) observations reveal diverse molecular gas morphologies and kinematics across the sample. Emission is detected in NGC~4636, NGC~5044, and NGC~5846, while NGC 5813 shows no significant detection within the data sensitivity. All maps have a spatial resolution of $\sim 0.6^{\prime \prime}$ , probing molecular gas in the central few kiloparsecs.
NGC 5044 exhibits strong, extended CO(2–1) emission within the central 3–5 kpc, with a clumpy morphology embedded in a diffuse component extending to $\sim$5 kpc. 
NGC 4636 contains more spatially confined molecular gas, appearing as discrete clouds within the inner 1–2 kpc and showing modest velocity gradients. NGC~5846 also hosts compact, irregular CO(2–1) emission near the nucleus, indicative of dynamically unsettled gas and consistent with low-level cooling from the intragroup medium.
No CO(2–1) emission is detected in NGC~5813, with a 3$\sigma$ upper limit of $\sim 0.3 \ Jy \ km \ s^{-1}$, suggesting any molecular gas is faint or highly fragmented.
Overall, the sample spans a range of cold gas properties, from extended diffuse emission in NGC~5044 to compact, disturbed structures in NGC~4636 and NGC~5846, and a non-detection in NGC~5813, likely reflecting different stages of AGN-regulated cooling and feedback.

\section{Analysis}\label{discussion}

\subsection{Dust Extinction Properties}

Dust in elliptical galaxies, both diffused in the hot gas and in cold clouds, could naturally 
result from a merger with a gas-rich galaxy.
Although this explanation is certainly correct in some cases, we believe that there is no supporting evidence for the merger hypothesis for the galaxies investigated in this work. For example in NGC~4636 the mean stellar age 10.3 Gyr \citep{sanchez06} is inconsistent with even a few percent contamination of younger stars from another galaxy. In general, far-IR emission does not correlate with Balmer line stellar ages \citep{temi05}. 
Also, the grain sputtering time is short compared to the time $\gta 10^8$ yr for a typical merging gas-rich dwarf galaxy to be destroyed by dynamical friction and tides.

The irregular clouds of dust in our target galaxies, the focus of this paper, seem to be in a highly transient state, orbiting in the galactic potential out to a few kpc, where the free-fall time is $\sim 10^7$ yr, and likely to be accreted by the SMBH via recurrent chaotic collisions (e.g., \citealt{gaspari17}).

In the following, we aim to evaluate the dust extinction properties in several localized regions. Additionally, we seek to determine whether significant variations in dust properties exist in specific galaxy locations.

Plots of A$_I$ versus A$_V$, with their total errors, for each of the identified high dust absorption regions are shown in the right panel of Figure~\ref{fg:ratio}
along with the extinction laws characterized by the ratio of total (V-band) extinction to selective extinction $R_V = A_V/E(B-V)$ = 2.3, 2.6, 3.1 and 5.0, where $E(B-V)$ is the color excess.
The adopted extinction curve uses the \texttt{G23} class from the \texttt{dust\_extinction} Python package,\footnote{\url{https://github.com/karllark/dust_extinction}} which provides the mean dust-extinction behavior as a function of $R_V$ at spectroscopic resolution from 912 \AA ~to 32 $\mu$m \citep{Gordon2023}. This represents an improvement over previous work, where all $R_V$ relations relied on combined spectroscopic and photometric data and/or did not span this full wavelength range.

Throughout most of the Milky Way ISM, the extinction curve is consistent with $R_V$ = 3.1, 
with higher values ($R_V \sim 5$) for dust in outer clouds of the Galaxy
\citep{mathis90, goud98}.
Low $R_V$ values are generally regarded as an indication of a relative low value
of large grains to small grains ratio, although grain composition can also affect $R_V$ values.

Left panel of Figure~\ref{fg:ratio} shows the location of the dust high-absorption regions 
in which the extinction has been calculated. 
Data for only four galaxies are available since 
HST data for NGC~5044 are available only in one optical band. 
These plots clearly show that while extinction values in some regions of the galaxies are consistent with the canonical Milky Way value of $R_V$=3.1, some areas of high absorption systematically exhibit lower $R_V$ values.

Through the analysis of extinction maps from a sample of 10 elliptical galaxies, \citet{goud94} found systematic evidence of low $R_V$ values in galaxies with large-scale dust lanes or rings. In these galaxies, $R_V$ ranges between 2.1 and 3.3, with an average lower than the canonical Galactic value of 3.1. This suggests that the dust grains responsible for the extinction of optical light are generally smaller than those in our Galaxy.
In contrast, ellipticals with dust irregularly distributed in patches and filaments exhibit {\it canonical} $R_V$=3.1 values -- e.g. 
{\it normal} grain sizes. 

The small sample of BGGs presented here reveals a different behavior. While the
ratio of $A_V/A_I$ measured in individual dust features is, on average, only slightly lower than the canonical value of $R_V$ = 3.1, many dust patches display significantly lower $R_V$ values.
Specifically, high dust absorption regions identified as region \#3 in NGC~4472, region \#2 in NGC~4636, regions \#6 and \#8 in NGC~5813, and regions \#4 and \#6 in NGC~5846 are best described by low values of $R_V \sim 2.1- 2.5$, rather than the canonical $R_V =3.1$.
The departures from the standard extinction law in these regions are statistically significant at the $\gtrsim3\sigma$ level. This indicates that the dust extinction curve in these zones is genuinely different from the canonical law, and it is very unlikely that the low $R_V$ values arise from random noise. Instead, they strongly suggest intrinsically different dust properties in these specific, high‑absorption regions.

The dust properties in NGC 5846, a massive central E0 galaxy, were previously studied by \citet{goud98} using optical imaging from the 3.5m NTT at ESO. Their extinction analysis showed Milky Way–like dust properties, and the dusty features they identified closely match the absorption structures revealed in our HST data. The key difference lies in resolution: the NTT/EMMI images (0.27$^{\prime\prime}$/pixel, with image quality constrained by a typical seeing of 1$^{\prime\prime}$ ) blur the main filament into a single structure, whereas the HST Planetary Camera (PSF FWHM ~0.05$^{\prime\prime}$) resolves it into multiple knots and fine filaments. This improved resolution allows detailed extinction studies and facilitates direct comparisons with ALMA CO cloud distributions.

The HST data presented here confirm that the majority of the central dust in NGC 5846 exhibits normal Galactic extinction characteristics. However, this study extends previous work by uncovering localized regions where the dust properties deviate from the canonical values.

\begin{figure*}[ht]
\begin{center}
    \includegraphics[width=15.47cm]{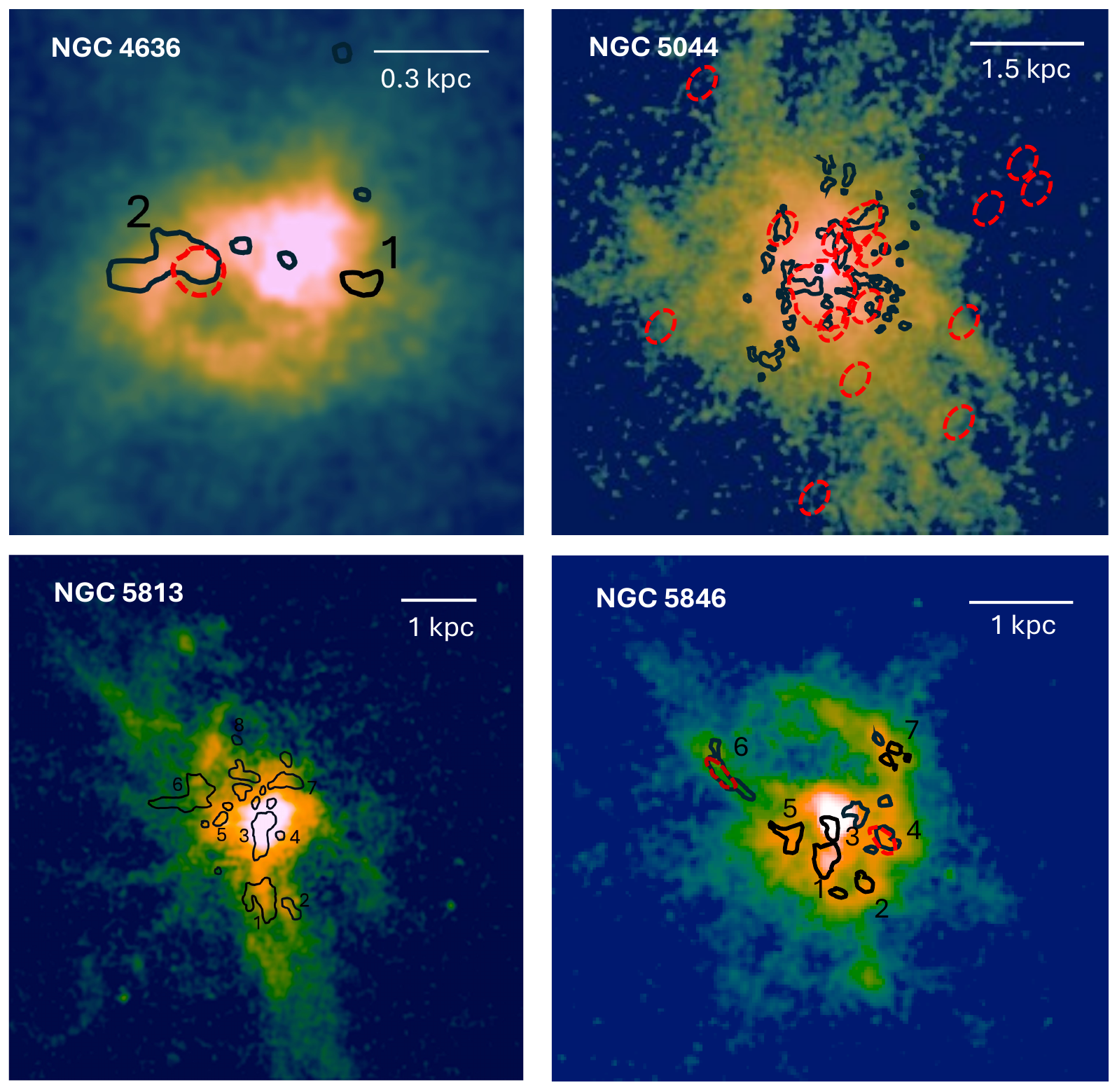}
  \caption{
MUSE H$\alpha$ images of NGC~4636, NGC~5044, NGC~5813, and NGC~5846, with regions of strong dust absorption outlined by black solid lines. Dashed red contours indicate ALMA CO-detected clouds in NGC~4636, NGC~5044, and NGC~5846 along with the labeled CO “cloud” ID numbers.  In NGC~4636 the highly extincted regions coincide with strong H$\alpha$ emission; dusty region \#2 overlaps with a CO cloud and shows enhanced $A_V/A_I$ relative to dusty region \#1. In NGC~5044 several resolved CO clouds lie within the central $\sim$2 kpc and overlap with regions of strongly extincted light as well as with the H$\alpha$ emission. Both NGC~5813 and NGC~5846 show a clear spatial correspondence between the knotty dust structures and the filamentary H$\alpha$ emission.
}
\label{fg:halpha}
\end{center}
\end{figure*}

\subsection{Regions with Distinct Dust Properties }

Figure \ref{fg:halpha} displays the H$\alpha$ emission maps for NGC~4636, NGC~5813, NGC~5846, and NGC~5044. Regions of strong dust absorption are outlined by black solid lines. ALMA CO-detected clouds are shown in dashed red contours along with the labeled CO “cloud” ID numbers.

ALMA observations of CO emission reveal two molecular cloud detections in NGC~4636, three in NGC~5846, and seventeen in NGC~5044 \citep{temi18}.
In NGC 5846, two of the three CO clouds (clouds associated with dust knots \#4 and \#6; hereafter \#4CO and \#6CO) are spatially resolved by the 12-meter array, with extents of 1.2$^{\prime\prime}$ ($\sim 160$ pc) and 2.9$^{\prime\prime}$ ($\sim 450$ pc), respectively \citep{temi18}. Cloud \#6CO in NGC~5846 is nearly perfectly aligned with a dust filament, while the other cloud in the field coincides with the small dust extinction features \#4 approximately 0.5$^{\prime\prime}$ wide ($\sim$60 pc). However, other similar dusty regions in the galaxy do not show detectable CO(2–1) emission.
Both CO clouds in NGC~4636 are unresolved in the ALMA data, indicating angular sizes $\leq 0.7^{\prime\prime}$ ($\sim$ 50 pc). Finally, one CO cloud in NGC~5846 and one in NGC~4636 lie outside the field of view of the MUSE H$\alpha$ emission map and are therefore not discussed here.
Many of the 17 detected CO clouds in NGC 5044 lie within $\sim 5^{\prime \prime}$ of the galaxy center. Only four are resolved by ALMA, three of which are in this central region. These central CO clouds align with strong dust absorption features, with the resolved ones encompassing multiple dust clumps, indicating a good correlation between dust and molecular gas within the central $\sim 5^{\prime \prime} \times 5^{\prime \prime}$.
Of the 17 clouds, 12 are located in regions of strong H$\alpha$ emission.
While molecular clouds in this region generally coincide with the brightest H$\alpha$ emission—suggesting a broad correlation—their spatial distribution does not strictly follow the H$\alpha$ morphology.

The limited number of detected CO(2–1) clouds (shown with dashed red contours) in NGC~4636 and NGC~5846 align well with both the warm ionized gas ($\sim10^4 K$) traced by H$\alpha$ emission and the cold dusty gas indicated by extinction. Where CO is detected, the molecular clouds appear to be co-spatial with both the dusty structures and the warm, line-emitting gas. 

As reported in the previous section, some of the highly extinct regions at the center of our sample galaxies
show distinctive dust properties.
Their derived $A_V/A_I$ ratios indicate that the dust properties are consistent with an extinction curve for $R_V \lesssim 2.5$.
It is of interest to investigate what set these regions apart from the remaining dust.\\
A clue emerges from the recent ALMA observations.
In NGC 4636 and NGC 5846, the three detected CO clouds coincide with regions of high starlight absorption that exhibit deviations from the canonical Galactic extinction law (see Figure \ref{fg:halpha}). 
This correlation suggests a physical connection between the presence of molecular gas and the altered dust extinction properties in these galaxies. 
Interestingly, several dust features revealed by HST in these galaxies do not show any associated molecular gas emission (see Figure~\ref{fg:dust1}). 
Because of the limited HST data available for NGC~5044, we cannot 
confirm that dust associated with CO has peculiar extinction characteristics for this galaxy.

Theoretical models and observational studies generally predict that dust associated with molecular gas—particularly in CO-bright regions—should exhibit higher total-to-selective extinction ratios, $R_V$, corresponding to flatter extinction curves. This expectation arises from enhanced grain growth in dense molecular environments, where shielding from UV radiation allows dust grains to coagulate and accrete icy mantles, increasing the average grain size \citep{draine03,jones96,ormel09,whittet01,chapman09}

However, in massive galaxies, this trend does not always hold. Deviations from typical grain growth are common in dynamic, energetic galactic centers in proximity of an AGN where strong radiation and shocks can disrupt grain growth by fragmenting large grains \citep[e.g.,][]{voit92, fabian2016}.
In addition, AGN-driven turbulence promotes large-scale mixing of the ISM, preventing the creation of dense, well-shielded environments required for grain growth and thereby sustaining a small-grain dominated dust population \citep{aoyama18,mattsson20}. 
Low $R_V$ values may also occur if the observed CO emission traces not only the dense, well-shielded cores of molecular clouds but also more diffuse or transitional molecular regions, where grain coagulation is less efficient and a higher fraction of small grains persists.

While the intense dynamics and energetics associated with AGNs are well recognized within the galaxy’s core, our observations reveal dusty regions and CO molecular clouds extending over several kiloparsecs, far beyond the immediate influence of the galactic nuclei. 
Thus, the observed peculiar dust extinction properties can be simply explained by the fact that the cooled dusty gas is relatively young, which indicates that the process of dust grain growth is still incomplete, resulting in environments dominated by small grains. However, since the formation of CO may involve molecular hydrogen in dusty regions associated with CO molecular emission, 
the dust presumably pre-exists the CO molecular clouds.

Furthermore, the long-standing view that the optical and near-infrared dust extinction curve is fully characterized by a single parameter, $R_V$, linked to grain size and composition, has recently been challenged by Gaia BP/RP spectra analysis of over 130 million stars. \citet{green24} show that extinction curves of the Milky Way cannot be described by $R_V$ alone, revealing how multiple independent components contribute to extinction variation which reflect heterogeneity in dust composition and local physical conditions of the ISM. A two-dimensional map of $R_V$  across the Milky Way \citep{zhang23} reveal substantial spatial variation in $R_V$, from molecular cloud scales to kiloparsec scales. A pronounced correlation emerges between $R_V$ and the locations of molecular clouds, with notably lower $R_V$ values observed within cloud interiors compared to their surroundings.

\subsection{Is there Excess Emission in the near-IR?}

Deviation from standard Galactic extinction law is usually regarded as an indication of dust properties, in terms of grain size distribution and/or chemical grain composition, that may differ from those of the ISM in the Milky Way. Here we want to investigate if the discovery of very low $R_V$ values only near the CO clouds may be due to I--band emission from
unusually hot, and possibly small, dust grains.

Near-IR emission from hot dust in hot plasma environments similar to 
those encountered in our sample galaxies is commonly observed \citep{kimura98,mann06}.
These observations indicate the presence of dust in the inner solar corona where a $10^{6}~K$ gas is present with high density $n_e \sim 10^8$\,cm$^{-3}$ \citep{hopkins18}. 
Dust dynamics, including formation and destruction and dust--gas interactions are responsible for the observed emission and dust features in the corona \citep{kimura98,kimura98b,kobayashi09}.
Although the hot gas density in the atmospheres of our galaxy sample is orders of magnitude lower, we are motivated to investigate whether the anomalous dust extinction properties in localized areas could be attributed to excess emission in the I-band.
While emission in the I-band
is rare, emission in the K-band is commonly observed \citep[][for a review]{mann06}.

We have conducted a detailed analysis of near-infrared observations from both 2MASS and \textit{Spitzer}-IRAC for galaxies with ALMA molecular gas detections, in order to identify localized infrared emission excesses at the positions of individual CO clouds.
Revealing small variation in surface brightness at the centers of these bright galaxies has proven to be difficult 
using a single band image. Contrast is much improved when an appropriate subtraction of J and K-band images is used.  
However, our investigation did not provide any evidence in support of near-IR excess emission at specific CO cloud locations.  

Although further investigation is necessary, with current data, we do not find evidence that 
the high $A_V/A_I$ ratio, observed 
in regions where dust is associated with CO clouds,
is due to I-band emission from hot dust rather than
deviation in dust properties from the ISM in the Galaxy.

\subsection{H$\alpha$ and CO Kinematics and Dust Correlation}\label{correlation}

\begin{figure}[ht]
\begin{center}
\includegraphics[width=8cm]{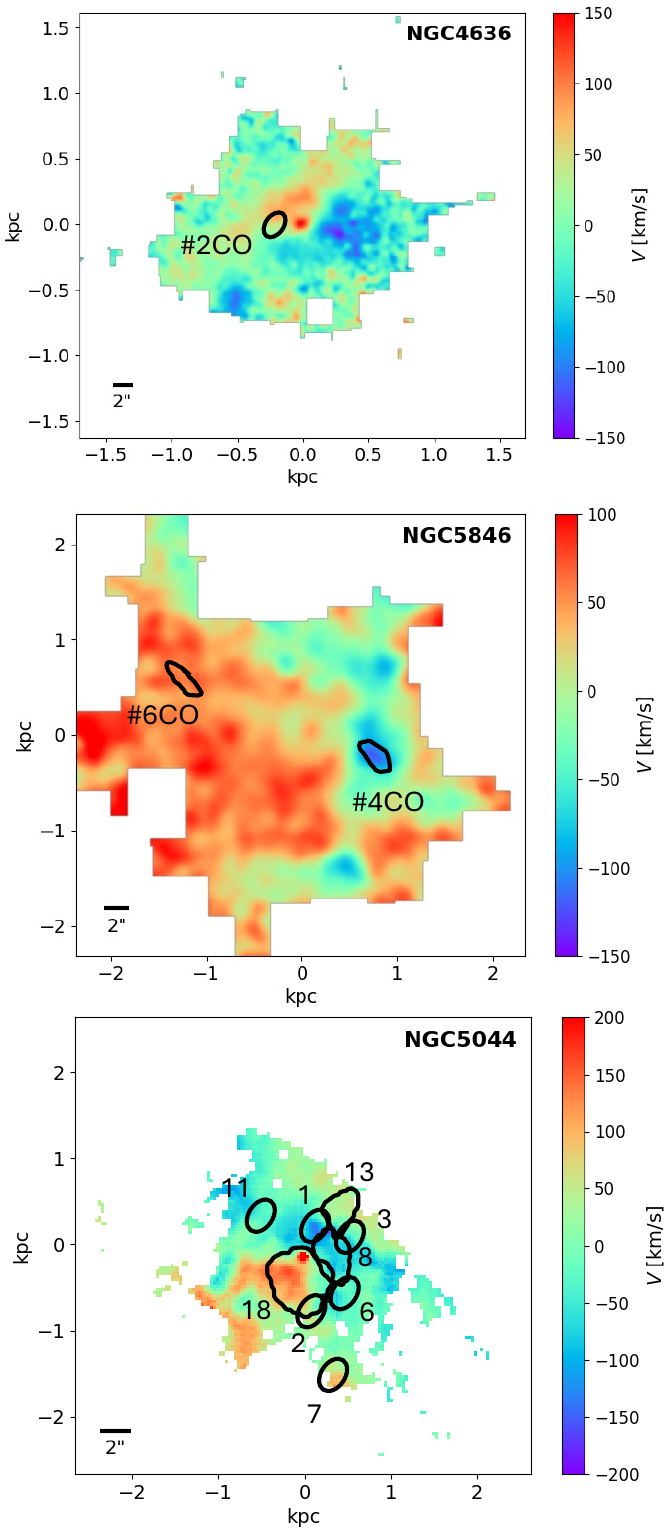}
  \caption{
H$\alpha$ radial velocity maps of NGC~4636, NGC~5846, and NGC~5044
  with ALMA CO detected clouds shown in black contours. 
}
\label{fg:velocity}
\end{center}
\end{figure}

Velocity maps of the H$\alpha$-emitting gas (Figure \ref{fg:velocity}) enable a comparison of the kinematic structure in regions spatially coincident with both molecular gas clouds and dust extinction. 
Detailed kinematic properties of the CO clouds within the galaxy sample are presented in \cite{temi18}. Here, we provide a summary of the key findings.\\
In NGC~5846, cloud \#6CO shows possible bimodality and a smooth velocity gradient (100–120 km s$^{-1}$), while \#4CO is unimodal with a velocity span across the cloud of $\Delta v = 25$ km s$^{-1}$ with an average of $\langle v \rangle$= -230 km s$^{-1}$. In NGC~4636, the cloud at dust knot \#2 aligns with an optical emission ridge and has $\langle v \rangle$= 200 km s$^{-1}$. Seventeen CO clouds are detected in NGC~5044, including several large (290–590 pc) or unresolved structures, with some outer clouds lacking associated dust. These data are summarized in Table~\ref{tbl:masshalpha_part2} and Table~\ref{tbl:masshalpha}.

The velocity maps of the H$\alpha$-emitting gas reveal coherent kinematic structures in regions spatially coincident with molecular gas clouds. In particular, clouds \#6CO and \#4CO in NGC~5846 lie within areas where the H$\alpha$-emitting gas displays opposite line-of-sight motions, yet both follow the global velocity gradient, suggesting a physical connection between the molecular and ionized components. Likewise, the CO cloud associated with dust knot \#2 in NGC~4636 shows a velocity pattern broadly consistent with that of the H$\alpha$-emitting gas. Although the velocity fields of the warm and cold gas phases exhibit strong overall agreement, their absolute velocities differ, indicating a significant offset between the two phases.

\begin{table}[t!]
  \begin{center}
   \caption{Kinematic properties of CO clouds and H$\alpha$ emission in {\bf NGC 5044}}
   \begin{tabular}{crrrr}
    \hline
    \hline
    CO  & $\langle V_{\rm H\alpha} \rangle$ & $\sigma_{\rm H\alpha}$ & $\langle V_{\rm CO} \rangle$ & $\sigma_{\rm CO}$ \\
    Cloud ID  & (km/s) & (km/s) & (km/s) & (km/s) \\
    \hline
   1  & -176 $\pm$ 41   & 86 $\pm$ 6   &  -559.6 $\pm$ 11.9 &  67.0 $\pm$ 10.7 \\
   2  & -81 $\pm$ 24   & 77 $\pm$ 9    &  -313.1 $\pm$ 9.5  &  36.6 $\pm$ 9.4 \\
   3  & -116 $\pm$ 47   & 71 $\pm$ 7    &  -274.4 $\pm$ 4.5  & 28.4 $\pm$ 4.4 \\
   6  &-106 $\pm$ 16   &83 $\pm$ 6    & -226.8 $\pm$ 4.3   & 20.4 $\pm$ 4.3  \\
   7  & -71 $\pm$ 12   & 71 $\pm$ 8   & -207.0 $\pm$ 6.9   &  31.1 $\pm$ 6.4 \\
   8  & -175 $\pm$ 46   & 93 $\pm$ 16  & -148.8 $\pm$ 8.2   & 76.2 $\pm$ 7.2 \\
   11 & -105 $\pm$ 27   & 79 $\pm$ 9   &-95.8 $\pm$ 6.0     & 41.0 $\pm$ 4.9  \\
   13 & -124 $\pm$ 55   & 75 $\pm$ 19  & -80.9 $\pm$ 6.9    & 43.1 $\pm$ 8.1 \\
   18 & 32 $\pm$ 15    & 82 $\pm$ 15  & 27.8 $\pm$ 1.7     & 37.4 $\pm$ 1.7 \\
    \hline
   \end{tabular}
    \label{tbl:masshalpha_part2}
\end{center}
  {\footnotesize
  \noindent\textbf{Notes.} 
  Cloud ID number refers to the CO clouds as identified in \citet{temi18}. Only CO clouds within the detected velocity field of the H$\alpha$ emitting gas are included.
  }
\end{table}

The H$\alpha$ velocity map of NGC 4636 shows a smooth gradient transition from blueshifted to redshifted velocities, indicating a coherent large-scale gas motion across the galaxy.\\
MUSE observations of NGC 5044 reveal a complex H$\alpha$ velocity field within the galaxy’s core. The velocity map on $\sim$ 1 kpc scale shows signs of a rotating disk-like structure near the nucleus with velocity amplitudes of $\pm$ 150 km/s, alongside more turbulent and disturbed motions in the surrounding filaments and extended emission regions. The velocity dispersion varies across the field, reflecting both dynamically rotating gas and more turbulent or shocked areas.  
The H$\alpha$ velocity field qualitatively agrees with the velocities of the individual clouds. Notably, the only redshifted cloud near the galaxy center coincides with the small region in the H$\alpha$ map that shows positive velocities within an area otherwise dominated by negative velocities. Again, their absolute velocities differ, revealing a notable velocity offset between the two phases.
Table~\ref{tbl:masshalpha_part2} summarizes the kinematic properties of the CO clouds and the corresponding H$\alpha$ emission regions in NGC 5044.

For the galaxies with both MUSE and 2 HST filter observations, we calculated the mass of the dust within the dust regions plotted in Figure~\ref{fg:ratio}
following the approach described in \citep{goud94}. We assumed a grain size between 0.005 and 0.22 $\mu$m \citep{draine84}, a specific grain mass density of 3 g/cm$^3$, a mixture of graphite and silicate (equal absorption from those), a grain size distribution proportional to a$^{-3.5}$ \citep{mathis77,goud94}.

Table~\ref{tbl:masshalpha} highlights the best-resolved and most representative structures and summarizes the kinematic properties of both the molecular gas and the $H\alpha$-emitting gas in regions corresponding to those identified in the extinction maps. Global $H\alpha$ properties for all galaxies are presented separately in Table~\ref{tbl:summary}. The table also reports the dust masses derived from V-band extinction measurements and the sizes of the “clouds” detected in dust absorption.

\begin{table}[t!]
  \begin{center}
   \caption{Physical and kinematic properties of regions with strong dust absorption.}
  \begin{tabular}{ccccc}
  \hline
   \multicolumn{4}{c}{{\bf NGC~4636}}\\
    \hline
    \hline
    Dust  & Size    & $M_\mathrm{d}$ & \hspace{-0.6cm }H$\alpha$ Flux\\ 
    Cloud ID    & (kpc$^2$) & (M$_\odot$) &  \hspace{-0.3cm }$10^{-16}$ erg/s/cm$^{-2}$\\
     \hline
    1  & 0.002 &  17.51 &   8.08 $\pm$   0.04   \\
    2  & 0.013 & 266.96 &  23.70 $\pm$   0.03   \\
    \\
    \hline
     \hline
    & $\langle V_{\rm H\alpha} \rangle$ & $\sigma_{\rm H\alpha}$ & $\langle V_{\rm CO} \rangle$ & $\sigma_{\rm CO}$ \\
     & (km\,s$^{-1}$) & (km\,s$^{-1}$) & (km\,s$^{-1}$) & (km\,s$^{-1}$) \\
      \hline
    \#2CO & 89 $\pm$ 8  & 49 $\pm$ 13 & 209 $\pm$ 4   & 25.8 $\pm$ 3.9 \\
   \\
    \multicolumn{4}{c}{{\bf NGC~5813}}\\
     \hline
     \hline
     Dust & Size    & $M_\mathrm{d}$ & \hspace{-0.6cm }H$\alpha$ Flux\\ 
    Cloud ID    & (kpc$^2$) & (M$_\odot$) & \hspace{-0.3cm }$10^{-16}$ erg/s/cm$^{-2}$ \\
    \hline
    1  &  0.067 & 2065.30 &  5.71 $\pm$   0.03 \\
    2  &  0.015 & 577.28 &  1.61 $\pm$   0.02  \\
    3  &  0.062 & 482.36 & 18.64 $\pm$   0.07  \\
    4  &  0.003 &  19.33 &  0.71 $\pm$   0.02  \\
    5  &  0.011 & 121.30 &  2.61 $\pm$   0.03  \\
    6  &  0.086 & 2242.58 & 6.75 $\pm$   0.08  \\
    7  &  0.080 & 591.54 & 13.03 $\pm$   0.07  \\
    8  &  0.003 &  67.86 &  0.46 $\pm$   0.02  \\
    \\
    \multicolumn{4}{c}{{\bf NGC~5846}}\\
     \hline
     \hline
      Dust & Size    & $M_\mathrm{d}$ & \hspace{-0.6cm }H$\alpha$ Flux\\ 
    Cloud ID    & (kpc$^2$) & (M$_\odot$) & \hspace{-0.3cm }$10^{-16}$ erg/s/cm$^{-2}$\\
    \hline
    1  &  0.002 &  31.42 &  3.19 $\pm$   0.03  \\
    2  &  0.010 & 197.30 & 17.90 $\pm$   0.05  \\
    3  &  0.063 & 572.12 &151.5 $\pm$   0.2    \\
    4  &  0.022 & 344.21 & 19.44 $\pm$   0.07  \\
    5  &  0.025 & 299.97 & 50.25 $\pm$   0.08 \\
    6  &  0.044 & 1912.73 &33.04 $\pm$   0.05  \\
    7  &  0.007 & 147.82 & 15.83 $\pm$  0.04 \\
    \\
     \hline
     \hline
     & $\langle V_{\rm H\alpha} \rangle$ & $\sigma_{\rm H\alpha}$ & $\langle V_{\rm CO} \rangle$ & $\sigma_{\rm CO}$ \\
     & (km\,s$^{-1}$) & (km\,s$^{-1}$) & (km\,s$^{-1}$) & (km\,s$^{-1}$) \\
      \hline
    \#4CO & -182 $\pm$ 12   & 81 $\pm$ 19 & -230 $\pm$ 2 & 23.3 $\pm$ 1.6 \\
    \#6CO & 62 $\pm$ 5    & 56 $\pm$ 18 & 110 $\pm$ 2 & 19.9 $\pm$ 1.6 \\
    \\
    \hline
  \end{tabular}
  \label{tbl:masshalpha}
  \end{center}
{\footnotesize
  \vspace{1ex}
  \noindent\textbf{Notes.} Column descriptions:
  (1) Dust Cloud ID number;
  (2) Size of the region with strong dust absorption in square kiloparsecs (kpc\textsuperscript{2});
  (3) Dust mass $M_d$ in solar masses ($M_\odot$);
  (4) H$\alpha$ flux measured in $10^{-16}$ erg\,s$^{-1}$\,cm$^{-2}$;
  For the dust clouds associated with CO clouds we report the following values: Average H$\alpha$ velocity in km\,s$^{-1}$;
  Velocity dispersion $\sigma$ of H$\alpha$ emission in km\,s$^{-1}$; Average CO velocity in km\,s$^{-1}$;
  Velocity dispersion $\sigma$ of CO emission in km\,s$^{-1}$.
  }
\end{table}

\begin{figure*}[t!]
\begin{center}
\includegraphics[width=15.6cm]{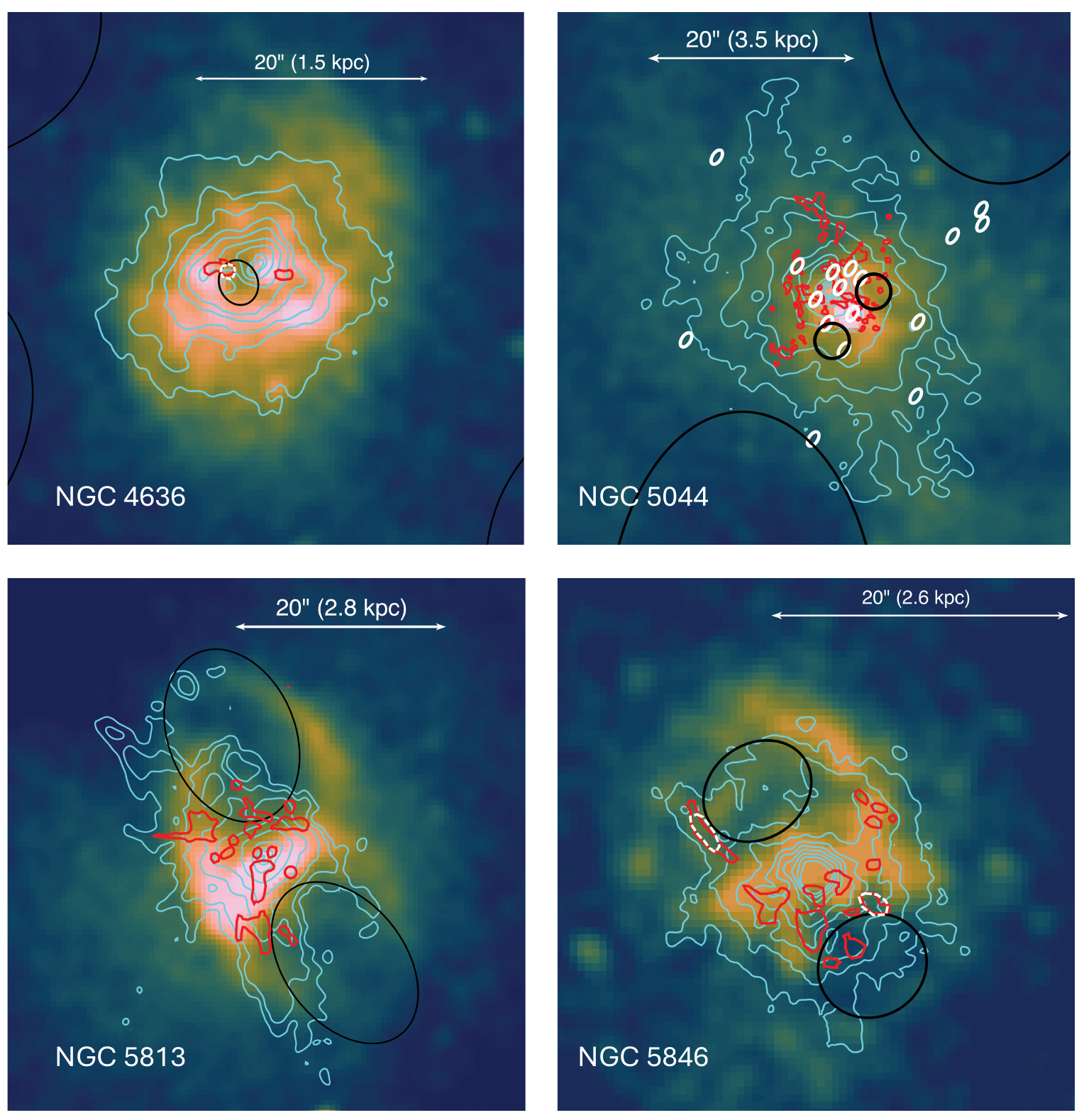}
  \caption{
  X-ray images of the central regions of NGC~4636, NGC~5846, NGC~5813, and NGC~5044, overlaid with H$\alpha$ emission contours in cyan. Regions of high dust extinction are shown with red contours, and molecular clouds detected in CO are outlined with white dashed ellipses. Inner X-ray cavities are marked by black ellipses, and in fields where present, outer cavity ridges near the image edges are traced with black lines.
}
\label{fg:cavities1}
\end{center}
\end{figure*}

\subsection{Role of X-ray Data in Interpreting Dust, CO and H$\alpha$ Emission}

X-ray data provide critical constraints for the interpretation of both dust and H$\alpha$ emission in galaxy groups. The close spatial (and kinematical) correlation between soft X-ray emission, warm ionized gas, and the cold molecular phase for NGC\,4636, NGC\,5846 and NGC\,5044 has previously been reported by \citet{temi18}. Similarly, \citet{olivares2022a} reported a correlation between H$\alpha$ luminosity and the cold molecular phase in a more extensive sample of brightest group galaxies. This alignment supports a condensation scenario (e.g., \citealt{gaspari11,gaspari12,valentini15,gaspari17,voit17a}), in which
compressive bulk motions and subsonic turbulence in the hot halo seed thermal instabilities that cascade into warm filaments and molecular clouds at local overdensity peaks. 

X-ray observations are also essential in revealing the footprints of AGN feedback in the hot atmosphere. Jets from the central AGN can displace the surrounding gas as they expand, creating low X-ray surface brightness structures filled by non-thermal plasma (e.g., \citealt{ubertosi21}). These structures, known as X-ray cavities, have been routinely observed in the hot atmospheres of galaxy groups (e.g., \citealt{dong10,eckert21}).\\
In this paper, we focus on the role of dust within the multiphase medium. In particular, we examine how dust-rich structures trace the spatial distribution of
molecular gas and H$\alpha$ filaments in the central regions of the galaxies, providing additional constraints on the formation and survival of cold gas in a hot atmosphere.\\
The group-centered elliptical galaxies analyzed in this investigation reveal a notable diversity in the spatial relationship between regions of high dust absorption and molecular gas (CO) emission. In NGC\,4636, and NGC\,5846, we find that all the CO-emitting clouds are associated with regions of strong dust absorption. However, most (lower extinction) dusty features have no detectable CO emission. 
In both cases the dust must be well mixed with cold gas and shielded from the ambient hot gas.

Figure \ref{fg:cavities1} shows X-ray images of the central regions of NGC~4636, NGC~5044, NGC~5813, and NGC~5846, with H$\alpha$ emission shown as cyan contours. Red contours mark areas of strong dust extinction, while white dashed ellipses outline CO-detected molecular clouds. Inner X-ray cavities are indicated by black ellipses, and where applicable, outer cavity ridges near the frame edges are delineated by black lines.\\
NGC~4636 is shown in Figure~\ref{fg:cavities1} over its central $\sim 40^{\prime\prime}$ of X‑ray emission. While \citet{baldi09} provide a comprehensive description of the larger‑scale X‑ray morphology, here we restrict our discussion to the smaller field of view covered by the MUSE and HST observations. In this region, the H$\alpha$-emitting gas displays a morphology that closely matches that of the hot X‑ray–emitting plasma.
A central depression in the H$\alpha$ emission corresponds to a cavity in the X-ray map (outlined by a black ellipse in the figure). The molecular cloud detected in this field of view is associated with a region of high dust extinction and is located along the rim of the central X-ray cavity.

In contrast, NGC\,5044 shows a somewhat more complex picture. Several unresolved molecular clouds (indicated by small white ellipses) are detected within the central few kpc, while a few larger cloud associations at the very center are resolved by ALMA 12 m array observations \citep[see Figure \ref{fg:halpha} and][]{david14, temi18}. Although most of the central CO clouds overlap dust-rich regions, some centrally located unresolved clouds do not show detectable dust absorption. Many unresolved CO features lie outside the field of view of the HST extinction map, preventing any analysis about their association with the dust. Notably, a subset of the central molecular clouds coincident with localized dusty knots appear to trace the edges of the two inner X-ray cavities (outlined by black ellipses).

NGC 5813 exhibits extended H$\alpha$ emission aligned with the axis of the X-ray cavities. Prominent dust features overlap the warm ionized gas and trace the rear edges of the cavities. In contrast to the other galaxies discussed above, the current ALMA observations do not detect CO(2–1) emission in NGC 5813. \citet{fujita24} derive an upper limit of $3.2 \times 10^{6}\ M_\odot$ for the molecular gas in the central 500 pc region, a value which classifies NGC 5813 as a relatively CO-poor system, similar to NGC 4636. 

In NGC 5846 the two CO clouds, outlined by white dashed contours, appear to align with the rims of the two identified cavities (black ellipses; \citet{machacek11}). Both the H$\alpha$-emitting gas and the dusty gas prominently trace the northern and southern rims of the X-ray emission, which outline the boundaries of the north and south cavities.

The presence of an extended reservoir of warm gas and dust-rich molecular clouds, often co-spatial with the rims of X-ray cavities, supports the scenario in which hot, low-entropy gas is uplifted and compressed by cavities and AGN outflows. This process can stimulate thermal instability and condensation, especially near cavity boundaries, leading to additional formation of cold clouds at the bubble edges \citep{brighenti02,gaspari12,voit17a, brighenti15,mcnamara16} on top of the quasi isotropic CCA rain \citep{gaspari17} generated by volume-filling turbulence.

Sputtering of grains is very effective in the hot gas (e.g. Tsai \& Mathews 1995, Draine 2011) and for temperatures $\gta 5\times 10^6$ K, typical of the six group-centered elliptical galaxies studied here, is primarily sensitive to the gas density. Our X-ray bright elliptical galaxies with extended cool gas have a typical density profile $n\approx 0.15 \,r_{\rm kpc}^{-0.7}$ cm$^{-3}$, where $r_{\rm kpc}$ is the radius in units of kpc \citep{lak18}. This results in a sputtering time $t_{\rm sputt} \approx 8\times 10^5 \,r_{\rm kpc}^{0.7}$ yr.
Thus, the hot gas is expected to be very dust-poor, with a typical dust-to-gas ratio $\delta \approx 10^{-4} - 10^{-6}$ \citep[e.g.][]{donahue93,tsai95,temi03, valentini15}. If cold gas originates from the thermal instability of the hot atmosphere of these central ellipticals, dust grains must therefore grow by accretion from dust seed, a mechanism which is effective in dense, low temperature gas \citep{dwek98, draine09}.

The typical timescale for dust growth by accretion in the densest and coldest clouds is $\sim 10^5 - 10^6$ yr \citep{dwek98}. In less dense or low-metallicity environments, where growth is slower due to lower collision rates between gas-phase metals and dust grains, the timescales can extend up to several tens or even hundreds of millions of years \citep{hirashita00,hirashita02,hirashita11,valentini15}. Thus, dust can effectively grow in dense gas with timescales comparable or shorter than the dynamical time at the cloud location (few kpc), which is an approximate measure of the age of the (dynamically transient) cold gas.
Additionally, variations in dust grain size distribution may occur depending on the age of the cold gas clouds.
This conclusion is supported by \citet{hirashita17}, who developed a dust evolution model for elliptical galaxies undergoing AGN feedback cycles and compared it with observational data, demonstrating that the dust-to-gas ratio is primarily governed by the dust growth timescale in the cold gas phase, and that the model range of accretion timescales ($\sim 10^6-10^8$ yr) effectively accounts for the observed dust and gas. This might apply to the systems studied in this work, being prominent examples of elliptical galaxies with several AGN feedback structures (especially NGC~5044, NGC~5813, and NGC~5846, which show clear X-ray cavities; Fig. \ref{fg:cavities1}).

A crucial question arises regarding the timescale for molecules to form in the cooled gas, and how this timescale compares to the other relevant processes, such as dust formation, gas cooling, and dynamical evolution, within the cold cloud environment. Molecular hydrogen requires dust to form rapidly, with timescale of $\approx 10^6$ yr, once the dust-to-gas ratio approaches values similar to the Galactic one \citep[e.g.][]{glover10, glover12}. CO is also expected to form with similar timescale. The cold gas in our galaxies is arranged in irregular and disturbed fashion (see Fig. \ref{fg:halpha}), so it is dynamically transient, with a likely age of the order of the dynamical time, $\approx 10^7$ yr. This very qualitative analysis shows that the hierarchy of timescales allows the dust and molecules to form in the cooled gas of the systems in our sample, although the uncertainties are large.

The dust growth process requires dust seeds. These can be provided by the small amount of dust surviving in the hot gas \citep[see][]{tsai95} or by the stellar mass loss \citep[e.g.][]{Mathews03}. Specifically, AGB stars are a likely a source of dust, through mass loss from evolved stellar populations. The stripped circumstellar dusty gas rapidly establishes pressure equilibrium with the hot interstellar gas and is ionized and (briefly) maintained at warm temperatures $T\sim 10^4 K $ by diffuse UV emission from hot post-AGB stars. 
It is expected that by $\sim 10^6$ yrs following its ejection from its parent star the dusty ionized gas is heated to $T\sim 10^7 K $ and thermally merges with the ambient hot gas, leaving the dust grains directly exposed to the hot interstellar gas \citep{mathews90, mathews03a, Mathews03}. After this short phase, the injected dust can be quickly sputtered away, generating dust-to-gas ratios of $10^{-5} - 10^{-6}$ \citep{tsai95, temi03}. However, in the central region of the galaxy dusty gas ejected by evolved stars can cool faster than the sputtering time, allowing a larger dust-to-gas ratio in the cooled gas \citep{Mathews03}. While the following evolution of this newly formed dusty clouds is uncertain, this mechanism can in principle supply dust seeds needed by the dust accretion process.

Yet another mechanism to (intermittently) feed dust in the hot atmosphere has been proposed by \citet{hirashita17}. In this scenario the AGN feedback periodically heats and ejects cold, dusty gas, effectively mixing it with the hot phase. This scenario, along with the close association of dusty molecular gas with the X-ray cavities in the hot atmospheres observed in NGC 4636, NGC 5846, NGC 5813 and NGC 5044 (Fig. \ref{fg:cavities1}), supports the idea that feedback from the central AGN has a primary role in the injection and survival of dust. 

\section{Summary and Conclusions}\label{conclusions}

The high spatial resolution of our multiphase dataset enables us to resolve and identify distinct dust and gas structures within the central kiloparsec of these galaxies, revealing signatures of a dynamically unsettled medium.
The results presented here indicate that only a subset of dusty structures in our sample of group-centered galaxies are associated with cold molecular gas. In NGC 4636 and NGC 5846, all the three detected CO clouds within the central kiloparsec align with regions of strong dust absorption. 
Furthermore, the consistent radial velocities of CO and nearby H$\alpha$ emission in these two galaxies further support a genuine physical association between dust, molecular and warm gas, rather than a chance alignment along the line of sight.
In contrast, only some of the CO clouds in NGC 5044 exhibit spatial correlation with dusty features. NGC 5813, albeit rich in dust absorption features, shows no detectable CO emission.

Most surprising is the discovery of unusually low values of the total-to-selective extinction ratio, $R_V$, localized near the CO clouds. This is counterintuitive: in regions rich in cold molecular gas, elevated $R_V$ values are typically expected due to grain growth and coagulation in dense environments. Instead, the observations reveal lower $R_V$  values, indicative of steeper extinction curves \citep[e.g.,][]{fitzpatrick07, Gordon2023}, suggesting that dust in these environments may be more heavily processed, fragmented, or otherwise evolutionarily distinct from dust in typical star-forming regions. Alternatively, clouds with low $R_V$ values may represent relatively young regions where dust growth remains incomplete, yielding a grain-size distribution dominated by small particles.

Several authors have contributed to the debate around the origin of the cold gas and dust in luminous elliptical galaxies with hot gas halos \citep[e.g.][and references within]{goud98, Mathews03, temi07a, temi07b, gaspari12, werner14, temi18, gaspari18, Eskenasy2024, olivares2022a, temi22, werner18}.
In all the galaxies studied here
chaotically arranged dusty fragments and filamentary structures are observed against the stellar background. These features appear to consist of cold gas in a highly transient state, orbiting within the galactic potential at distances of several kpc, where the free fall time is $\sim 10^7$ years.
Given their size and associated dust mass, it is highly unlikely that these relatively large, optically obscuring clouds are infalling material recently ejected by stellar processes.
Furthermore, the presence of molecular clouds in the form of off-center, orbiting structures argues against simple, axisymmetric inflow models and instead supports a more chaotic, clumpy condensation process in the hot X-ray-emitting halo. This is consistent with models of CCA/precipitation, in which radiative cooling from the hot gas halo leads to the intermittent formation of cold clouds, some of which may rain down toward the galaxy center and fuel SMBH accretion \citep[][]{gaspari13,Gaspari20,voit17}.


Hereafter we summarize the results from this study:

{\bf 1} -- We observe a clear spatial variation in dust extinction curves that correlates with the association of dust with molecular gas. Dust spatially coincident with molecular gas shows significantly steeper extinction curves and lower $R_V$  values ($\sim$ 2.0 - 2.5), indicative of a grain size distribution skewed toward smaller particles. In contrast, dust found outside CO-emitting molecular regions exhibits extinction characteristics similar to those of the diffuse interstellar medium in the Milky Way, though the total-to-selective extinction ratio spans a range of approximately 
 $R_V \sim 2.5 - 3.1$.

{\bf 2} -- Where CO and dust coexist, observations indicate a regime dominated by dynamic condensation–fragmentation processes, driven by turbulence and cooling. In these environments, dust grain shattering might be efficient, resulting in relatively low values of the total-to-selective extinction ratio $R_V$.
Alternatively, clouds exhibiting low values may represent relatively young environments where dust growth has not yet fully progressed, resulting in a grain-size distribution still dominated by small particles.

{\bf 3} -- Regions exhibiting dust absorption without associated CO may represent an evolutionary phase where initially CO-rich and dusty clouds have undergone molecular dissociation, leaving behind dust grains exposed to energetic processes such as shattering and sputtering in the ambient medium. Measured $R_V$ values within these dust-only regions lie in the range 2.5 to 3.1. This range suggests a trend where dust, once embedded in CO-rich environments with high $R_V$, is subsequently evolving towards lower $R_V$ values consistent with grain fragmentation and erosion. While speculative, this scenario aligns qualitatively with modeled evolutionary timescales discussed in this study.

{\bf 4} -- We find a strong spatial correspondence between dust extinction features and warm ionized gas traced by H$\alpha$ emission. Across multiple systems (NGC 4636, NGC 5813, NGC 5846), this correlation extends from compact CO clouds to diffuse, filamentary structures on scales of 20$^{\prime \prime}$–30$^{\prime \prime}$, indicating that filamentary and diffuse dust structures align closely with H$\alpha$ emission and are co-spatial over a range of morphologies and spatial scales. Such spatial alignment is expected if turbulent mixing or thermal conduction acts efficiently at the interfaces between hot and cold phases, consistent with the CCA multiphase condensation framework.

{\bf 5} -- Kinematic analyses reveal that both the molecular gas and the warm H$\alpha$-emitting gas display coherent velocity structures, alongside evidence of rotation and turbulence. However, some clouds exhibit complex brightness and velocity patterns, while some dust-rich regions lack corresponding molecular detections, indicating localized differences in the formation or survival of molecular clouds within these group-centered elliptical galaxies.

{\bf 6} -- The partial alignment of dusty, molecular, and ionized structures with X-ray cavities suggests a close connection with feedback, whereby rising bubbles and associated compressive motions amplify local thermal instabilities and trigger condensation on top of the volume-filling turbulent condensation.
While CO clouds are always associated with dusty features, many dust structures lack CO, implying additional reservoirs of diffuse cold gas below current detection limits. The survival and replenishment of dust can be explained by rapid grain growth in cold dense gas, occurring on timescales comparable to or shorter than both dust destruction and cloud dynamical lifetimes. Molecular gas forms in parallel once sufficient dust is present, reinforcing the interpretation of an in-situ growth process regulated by AGN feedback cycles.


\section{acknowledgements}
This paper employs a list of Chandra datasets, obtained by the Chandra X-ray Observatory, contained in the Chandra Data Collection (CDC) 507~\dataset[doi:10.25574/cdc.507]{https://doi.org/10.25574/cdc.507}. HST data is available at MAST: \dataset[doi:10.17909/7exc-yj68]{\doi{10.17909/7exc-yj68}}.
This paper makes use of the following ALMA data: ADS/JAO.ALMA\#2015.1.00860.S and ADS/JAO.ALMA\#2011.0.00735.SSB. ALMA is a partnership of the ESO (representing its member states), NSF (USA), and NINS (Japan), together with the NRC (Canada), MOST and ASIAA (Taiwan), and KASI (Republic of Korea), in cooperation with the Republic of Chile. The Joint ALMA Observatory is operated by the ESO, AUI/NRAO, and NAOJ.The National Radio Astronomy Observatory and Green Bank Observatory are facilities of the U.S. National Science Foundation operated under cooperative agreement by Associated Universities, Inc.\\
Based on observations made with the NASA/ESA Hubble Space Telescope, and obtained from the Hubble Legacy Archive, which is a collaboration between the Space Telescope Science Institute (STScI/NASA), the Space Telescope European Coordinating Facility (ST-ECF/ESA) and the Canadian Astronomy Data Centre (CADC/NRC/CSA). STSDAS is a product of the Space Telescope Science Institute, which is operated by AURA for NASA.
PT, KF, PM, AM and AB, acknowledge support from NASA’s NNH22ZDA001N Astrophysics Data and Analysis Program under award 24-ADAP24-0011. 
FU, MyG acknowledge support from the research project PRIN 2022 ``AGN-sCAN: zooming-in on the AGN-galaxy connection since the cosmic noon", contract 2022JZJBHM\_002 -- CUP J53D23001610006. MaG acknowledges support from the ERC Consolidator Grant \textit{BlackHoleWeather} (101086804). VO acknowledges support from the DICYT ESO-Chile Comite Mixto PS 1757, Fondecyt Regular 1251702.\\
We remember Prof. William, G. Mathews with appreciation for his valuable contributions to the conception and early development of this study.

 \bibliography{Dust_v4.bib}

\end{document}